\documentclass[a4paper,10pt,superscriptaddress,pra,showpacs,twocolumn]{revtex4}
\usepackage{amssymb}
\usepackage{amsmath}
\usepackage{float}
\usepackage{graphicx}
\usepackage{pstricks}
\usepackage{pst-grad}
\usepackage{pst-node}
\usepackage{pst-coil}
\usepackage{multido}
\usepackage{pst-plot}
\usepackage{psfrag}
\usepackage{float}
\usepackage{multirow}

\newcommand{\bra}[1]{\ensuremath{\langle#1|}}
\newcommand{\ket}[1]{\ensuremath{|#1\rangle}}

\newcommand{\tr}{\mathrm{tr}}

\newcommand{\mc}{\mathcal}

\renewcommand{\Re}{\textrm{Re}}

\newcommand{\diag}{\textrm{diag}}

\newcommand{\PP}{\mathcal  P}
\newcommand{\QQ}{\mathcal Q}
\newcommand{\DD}{\mathfrak D}
\newcommand{\ZZ}{\mathbf Z}
\newcommand{\XX}{\mathbf X}

\usepackage[latin1]{inputenc}
\usepackage{amsfonts}
\usepackage{amsmath}
\usepackage{amssymb}
\SpecialCoor

\psset{xunit=1,yunit=.6}
\psset{linearc=.01,fillstyle=solid}

\definecolor{LightOrange}{cmyk}{0,0.2,0.4,0}%
\definecolor{DarkOrange}{cmyk}{0,0.2,0.8,0}%

\definecolor{MeasureCol1}{rgb}{.4,0.5,1}
\definecolor{MeasureCol2}{rgb}{.4,1,1}
\definecolor{SymplecticCol1}{rgb}{.9 1 .9}
\definecolor{SymplecticCol2}{rgb}{.4 1 .6}
\definecolor{ThermalCol1}{rgb}{1 0 0}
\definecolor{ThermalCol2}{rgb}{1 1 .4}

\def\symplectic#1(#2){%
\pspolygon[fillcolor=cyan,linearc=.2, fillstyle=gradient,gradmidpoint=1,gradangle=45,gradbegin=SymplecticCol1,gradend=SymplecticCol2](0,-.5)(1,-.5)(! 1 -.5 #2 add)(! 0 -.5 #2 add)%
	\rput(!.5 -.5 .5 #2 mul add){#1}%
	}

\def\thermalTm(#1,#2){%
\psset{fillstyle=gradient,gradmidpoint=1,gradangle=0,gradbegin=ThermalCol1,gradend=ThermalCol2}
	\rput(!#1 .5 add #2){\rotateleft{\pscustom{\parabola(-.1,0)(0,.2)%
	\closepath
	}}}
}

\def\thermalFm(#1,#2){%
	\psset{fillstyle=gradient,gradmidpoint=1,gradangle=45,gradbegin=ThermalCol1,gradend=ThermalCol2}
	\rput(!#1 .5 add #2){\rotateleft{\pscustom{\parabola(-.1,0)(0,.2)%
	\closepath
	}}}
}

\def\gap{.08 }
\def\measure(#1,#2,#3){%
\psellipticarc[showpoints=true,fillstyle=gradient,gradmidpoint=1,gradangle=45,gradbegin=MeasureCol1,gradend=MeasureCol2](! #1 -.5 #2 #3 2 div add add)(!1 #3 2 div \gap sub){-90}{90}
	\psline(!#1 #2 -.5 add \gap add)(! #1 -.5 #2 #3 add add \gap sub)
	}

\begin{document}
\title{Measurement of damping and temperature:\\Precision bounds in Gaussian dissipative channels}

\author{Alex Monras}

\affiliation{Dipartimento di Matematica e Informatica,
Universit\`{a} degli Studi di Salerno, CNR-SPIN,
CNISM Unit\`{a} di Salerno, and INFN Sezione di Napoli,
Gruppo collegato di Salerno,Via Ponte don Melillo,
I-84084 Fisciano (SA), Italy}

\author{Fabrizio Illuminati}

\affiliation{Dipartimento di Matematica e Informatica,
Universit\`{a} degli Studi di Salerno; CNR-SPIN;
CNISM Unit\`{a} di Salerno; and INFN Sezione di Napoli,
Gruppo collegato di Salerno,Via Ponte don Melillo,
I-84084 Fisciano (SA), Italy}

\affiliation{Corresponding author. Electronic address:
illuminati@sa.infn.it}

\date{October 1, 2010}

\begin{abstract}
We present a comprehensive analysis of the performance of different classes of Gaussian states in the estimation of Gaussian phase-insensitive dissipative channels. In particular, we investigate the optimal estimation of the damping constant and reservoir temperature. We show that, for two-mode squeezed vacuum probe states, the quantum-limited accuracy of both parameters can be achieved simultaneously. Moreover, we show that for both parameters two-mode squeezed vacuum states are more efficient than either coherent, thermal or single-mode squeezed states. This suggests that at high energy regimes two-mode squeezed vacuum states are optimal within the Gaussian setup. This optimality result indicates a stronger form of compatibility for the estimation of the two parameters. Indeed, not only the minimum variance can be achieved at fixed probe states, but also the optimal state is common to both parameters. Additionally, we explore numerically the performance of non-Gaussian states for particular parameter values to find that maximally entangled states within $d$-dimensional cutoff subspaces [$d\leq6$] perform better than any randomly sampled states with similar energy. However, we also find that states with very similar performance and energy exist with much less entanglement than the maximally entangled ones.

\end{abstract}


\maketitle

\section{Introduction}
Decoherence lies at the core of all difficulties in implementing quantum information technologies. It degrades information being transmitted, stored, and processed, in an irreversible way. All these processes can be thought of different kinds of quantum channels, decoherence inevitably affecting all of them. We are interested in assessing the deviation from ideality in Gaussian channels, and the precision attainable when tested with Gaussian resources. Namely, we consider a dissipative thermal bath with mean photon number $N$ and damping constant $\gamma$, and explore how well different Gaussian resources perform in identifying these parameters. Our question is immediately relevant to the field of quantum information, since assessing the deviation from the \emph{identity channel}, \emph{i.e.}, the ideal information transmitter, is the principal requirement to implement large scale quantum communication. The need for an efficient characterization of dissipation in continuous variable systems is becoming a requisite for a number of quantum information tasks, such as quantum repeaters~\cite{azuma_optimal_2009, azuma_tight_2010} or quantum memories~\cite{kozhekin_quantum_2000,htet_characterization_2008, jensen_quantum_2010}, among others. The burden of dissipation is also hindering advances in cavity QED~\cite{brune_process_2008, deleglise_reconstruction_2008} and superconducting quantum circuits~\cite{wang_measurement_2008}.

On the other hand, measuring decoherence is not only relevant for quantum information technology. In several contexts, decoherence can be related to physical quantities of practical interest, \emph{e.g.}, photon loss is strongly related to impurity doping concentration in semiconductor lasers~\cite{hunsperger_photon_1969}. Nonlinear magneto-optical effects~\cite{budker_resonant_2002} can be understood as photon loss, an effect with several technological applications such as low-field magnetometry and gas density measurements~\cite{budker_nonlinear_1998, budker_sensitive_2000, shah_subpicotesla_2007}. Additionally, photon-photon scattering in vacuum is still an unobserved prediction of both quantum electrodynamics and non-standard models of elementary particles~\cite{tommasini_precision_2009}. These are only a few among the several applications that involve the precise determination of losses in dispersive media. For this reason, we will pose our problem and formulate our results in a general theoretic formalism, in order to keep our results as general as possible.

We address the problem of estimating the parameters of a Gaussian channel describing the dynamics of a bosonic mode $a$, coupled with strength $\gamma$ to a thermal reservoir with mean photon number $N$. In the interaction picture, and within the Markovian approximation (at any time the mode and the bath remain unentangled), the completely positive dynamics and the action on the mode $a$ is described by the superoperator
\begin{equation}\label{eq:channel}
	\mc S(\gamma,N)=\exp\frac{\gamma}{2}\left(NL[a^\dagger]+(N+1)L[a]\right),
\end{equation}
with $L[o]\rho=2o\rho o^\dagger-o^\dagger o \rho-\rho o^\dagger o$. For convenience we arrange the channel parameters in the two-dimensional vector $\theta=(\gamma,N)$ and let $\hat\theta$ be the estimator, corresponding to the outcome of the final measurement.

\begin{figure}[b]
\begin{center}
\includegraphics[width=.5\textwidth]{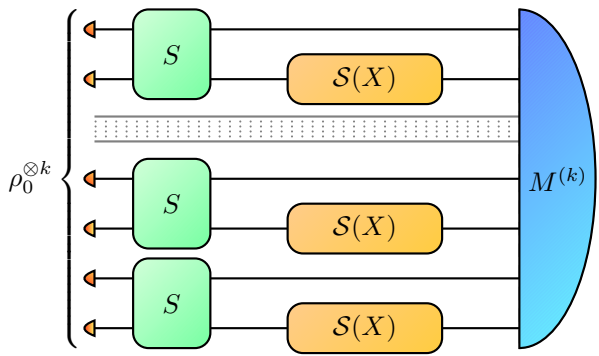}
\end{center}
\caption{\label{fig:strategy}
(Color online)~Scheme for measuring the values of parameters $\gamma$ and $N$. $N$ copies of a bipartite state $\rho$ are independently prepared, and acted upon by $k$ instances of the unknown channel $\mc I\otimes\mc S(\theta)$. A collective measurement $\Lambda$ is finally performed.}
\end{figure}

Previous related work in the literature addresses the problem of estimating a state within a Gaussian family in several different situations. Among others we should mention the works of Yuen \& Lax~\cite{yuen_multiple-parameter_1973}, Helstrom~\cite{helstrom_quantum_1976}, Holevo~\cite{holevo_probabilistic_1982}, Hayashi~\cite{hayashi_asymptotic_2008}, Adesso \& Chiribella~\cite{adesso_chiribella} and Hayashi \& Matsumoto~\cite{hayashi_quantum_2009}. However, all of these works focus on estimating some of all possible parameters of a Gaussian state. Most of them focus on estimating either displacement or temperature for fixed degrees of squeezing, while others consider the degree of squeezing in a vacuum state. No work exists, to the best of our knowledge, that addresses the problem of estimating \emph{all} parameters of a Gaussian state. If such a work would exist, the problem of estimating a Gaussian channel with Gaussian probe states would reduce to a subproblem of the former. However, the lack of such a general result demands a dedicated solution.

The setup that we consider is quite generic. We allow for \emph{i)} extending the channel of interest to include an ancillary mode $b$, unaffected by the channel (\emph{i.e.}, identity superoperator $\mc I$), obtaining the channel $\mc S^\star=\mc S\otimes\mc I$, \emph{ii)} choosing any bipartite Gaussian state $\rho_0$, of which $k$ copies will be sent through the channel and \emph{iii)} performing a generalized measurement $M^{(k)}$ [characterized by a POVM $\{M^{(k)}_{\hat\theta}\}$, $M^{(k)}_{\hat\theta}\geq0$, $\int d{\hat\theta}\,M^{(k)}_{\hat\theta}=\openone$], on the collective state $\rho_\theta^{(k)}=(\mc S^\star \rho_0)^{\otimes k}$. This allows for a rather general scheme, which tests the channel with independent and identically prepared probes, while the generalized collective measurement may include arbitrary quantum transformations applied to $\rho_\theta^{(k)}$ prior to the measurement [see Fig.~\ref{fig:strategy}].


As a quantifier of the quality of the estimate one can use the covariance matrix
\begin{equation}
	V_\theta(M)=\int d\hat\theta\,(\hat\theta-\theta)(\hat\theta-\theta)^\top\tr[\rho_\theta M_{\hat\theta}],
\end{equation}
which may depend on the chosen measurement $M=\{M_{\hat\theta}\}$ and the particular channel being tested, $\theta$.
Alternatively, as is customary in standard statistical inference, we can relate any error cost function $\ell(\theta,\hat\theta)$ to the covariance matrix by performing a Taylor expansion of $\hat\theta$ around $\theta$, and defining $G_\theta=\frac{1}{2}\partial^2 \ell(\theta,\hat\theta)|_{\hat\theta=\theta}$ as the Hessian of $\ell/2$, thus
\begin{equation}\label{eq:objective}
	\langle\ell\rangle=\tr[G_\theta V_\theta(M)]+o(\hat\theta-\theta)^2.
\end{equation}
More generally, an arbitrary positive semidefinite weight matrix $G_\theta$ can be defined to account for the relevance assigned to each parameter. Two extreme cases, where one only cares about one or the other parameter can be accounted for by the choices $G_\gamma=\diag(1,0)$ and $G_N=\diag(0,1)$. This approach allows also to define strategies where one is only interested in a particular linear combination of the parameters $X=x^\mu\theta_\mu$, by setting $G_X=XX^\top$. With these considerations, and given a large number $k$ of copies of the fixed probe state $\rho_0$, one can ask what is the smallest possible value of $\langle\ell\rangle=\lim_{k\rightarrow\infty} k\,\tr[G_\theta\,V_\theta(M^{(k)})]$ that is allowed by the laws of Quantum Mechanics.

The main question we wish to answer is the following: To what extent can one optimize expression~\eqref{eq:objective} by using different Gaussian resources? More precisely, we will compare the performance of thermal, coherent, single-mode squeezed and two-mode squeezed vacuum in estimating the damping $\gamma$ and the temperature $N$ at different parameter regimes, comparing the performance of each probe state at equal amount of energy input to the channel, $n=\tr[\rho_0 a^\dagger a]$. In order to elucidate the role that each one of these resources plays in the estimation problem, we will make some simplifying assumptions. Namely, we will not consider the combination of different resources, e.g. two-mode squeezing and displacement. 

A priori it would seem that the problem may have several different variants depending on the chosen cost function (or $G$ matrix). We will see, however, that some general statements can be made. Anticipating the
results that will be presented in the present work, we will prove that:
\begin{enumerate}
	\item Choosing a two-mode squeezed vacuum input state $\rho_0$, the parameters $\gamma$ and $N$ can be optimally estimated \emph{simultaneously}. That is, no compromise is required in the optimization of $V_\theta(M)$. This holds true even when the optimal measurements for $\gamma$ and $N$ do not commute.
	\item For both parameters $\gamma$ and $N$, and at any given energy, two-mode squeezed states always outperform any other class of Gaussian states.
\end{enumerate}
The combination of these two statements unveals a strong compatibility between the problem of estimating the damping $\gamma$ and the temperature $N$ in the Gaussian setting. In Section~\ref{sec:Holevo} we develop our approach and derive the first result of our paper, namely, that optimal precision with two-mode squeezed states can be attained simultaneously for both parameters, thus allowing to divide the problem of computing precision bounds into two independent problems. Sections \ref{sec:gamma} and \ref{sec:N} explore the precision bounds for estimating $\gamma$ and $N$ individually, comparing the performances of different Gaussian resources, focusing especially on some physically relevant regimes of the parameters. In Section~\ref{sec:nonG} we move on to explore numerically the non-Gaussian arena and compare the relative
performances with respect to Gaussian probes. Section~\ref{sec:discuss} concludes the paper with a discussion and an overview of the obtained results. Details of the technical proofs are provided in two appendices.

\section{The ultimate quantum limits}\label{sec:Holevo}
The Heisenberg relations place a fundamental limit on the precision with which one can measure any given observable. When it comes to quantities not associated to an observable, as is our case, it is necessary to resort to quantum estimation theory, which studies the fundamental quantum mechanical limits to the precision of measurements in a variety of situations. A first lower bound can be obtained from Helstrom's Fisher information matrix~\cite{helstrom_quantum_1976, holevo_probabilistic_1982},
\begin{equation}\label{eq:CRbound}
	V_\theta(M)\geq J(\theta)^{-1}~~~~\forall M
\end{equation}
where $J(\theta)=\Re\,\tr[\rho_\theta\,\Lambda \,\Lambda^\top]$ is defined as the covariance matrix of the \emph{symmetric logarithmic derivatives} (SLD) $\Lambda_\mu$ fulfilling $\partial\rho_\theta/\partial\theta^\mu=\Lambda_\mu\circ\rho$, with $A\circ B=(AB+BA)/2$. The inherent non-commutativity of Quantum Mechanics forbids, in general, to attain this inequality when the problem is multi-parametric, as in our case. Optimizing the measurement for one parameter will in general compromise the measurement precision on the others. When considering single-parameter estimation problems, it is well known that local adaptive measurements attain Eq.~\eqref{eq:CRbound}~\cite{gill_state_2000, hayashi_statistical_2003}. However, even if the optimal measurements for both parameters do not commute, it may still be possible to devise a measurement strategy to attain simultaneously both bounds. Recent progress in the theory of Local Asymptotic Normality for quantum states~\cite{gu_local_2007, kahn_local_2009} suggests that equality in Eq.~\eqref{eq:CRbound} is asymptotically attainable if and only if~\cite{guta_unpub},
\begin{equation}\label{EQ:COMMUTE}
	\tr[\rho_\theta\,[\Lambda_\mu,\Lambda_\nu]]=0.
\end{equation}

The SLD's for our problem, as well as in more general contexts, were obtained by the authors in~\cite{monras_information_2009}. In Appendix~\ref{sec:commute} we prove Eq.~\eqref{EQ:COMMUTE} for the case of two-mode squeezed vacuum probe states. The implications of Eq.~\eqref{EQ:COMMUTE} are two-fold. On one hand, it allows to prove that asymptotically, the estimation problems for both parameters become independent, so that they can be analyzed separately. This will be the subject of the two next Sections. On the other hand, we will show that two-mode squeezed states form the optimal Gaussian class of states. As a consequence, we will prove the existence of and provide the explicit expression of precision bounds for the simultaneous estimation of $\gamma$ and $N$ for error cost functions that have diagonal $G$ matrices.


The single-parameter precision is quantified by the asymptotic standard deviation,
\begin{equation}
	\Delta\theta=\sqrt{\lim_{k\rightarrow\infty}k\int d{\hat\theta} \,(\hat\theta-\theta)^2\tr[\rho_\theta^{(k)} M_{\hat\theta}^{(k)}]},
\end{equation}
which is bounded by the quantum Fisher information. The latter will be generically denoted as $J_\gamma$ or $J_N$ depending whether we are considering the yield for parameter $\gamma$ or $N$, respectively. When no confusion arises, we will omit the subscript. When we are referring to a particular yield of QFI, specific to a given class of states, we will denote it with the corresponding subscript $J_\textrm{coh.}$, $J_\textrm{th.}$, $J_\textrm{sq.}$ and $J_{\textrm{2-m}}$ for coherent, thermal, single-mode squeezed and two-mode squeezed vacuum states, respectively.
Precision bounds are then given by
\begin{eqnarray}
	\Delta \theta&\geq&\frac{1}{\sqrt{J_\theta}}
\end{eqnarray}
We will call $J_\theta$ the \emph{yield}, or \emph{performance}.

The obtained results should be interpreted in the following way. The two single-parameter problems have, as optimal observables the corresponding SLD's~\cite{braunstein_statistical_1994}, which can be explicitly computed from~\cite{monras_information_2009} and are quadratic in the creation and annihilation operators. In the multiparametric case, with diagonal $G$ matrices, the bounds obtained in the following two sections provide all necessary quantities needed to determine the asymptotic error cost. The optimal measurement, however, will require a general collective measurement, which is likely to be beyond the technical capabilities of present-day technology. We will focus on the theoretical attainable precision and not discuss the details of the implementation of the optimal observables. This will, nevertheless, provide a means to gauge the efficiency of more applied studies such as tomographic~\cite{bellomo_reconstruction_2009}, single-mode Gaussian~\cite{monras_optimal_2007} and non-Gaussian~\cite{adesso_optimal_2009}, or entanglement-assisted schemes~\cite{venzl_quantum_2007}.

In order to determine the precision attainable with different Gaussian resources we consider single-mode probes parameterized as
\begin{equation}
	\rho_0=D({\bf d}_0)S(r_0)\rho_{\nu_0} S^\dagger(r_0)D^\dagger({\bf d}_0).
\end{equation}
where $\rho_\nu\propto \left(\frac{\nu-1/2}{\nu+1/2}\right)^{a^\dagger a}$, is a single-mode thermal state, $D({\bf d})=\exp i(d_2Q-d_1P)$ [${\bf d}=(d_1,d_2)$] and $S(r)=\exp\frac{1}{2}(r a^{\dagger}{}^2-r^* a^2)$. The first moments and covariance matrix of the single-mode states ($\Sigma_0$) are given by ${\bf d}_0$ and
\begin{eqnarray}\label{eq:CM.sm}
	\Sigma_0&=&\frac{\nu_0}2\left(\begin{array}{cc}e^{2r_0} & 0 \\0 & e^{-2r_0}\end{array}\right),
\end{eqnarray}
while the energy in the probe is given by
\begin{equation}
	n=\frac{\nu_0\cosh 2r_0+|{\bf d}_0|^2-1}2.
\end{equation}

Using the results of~\cite{monras_information_2009} it is easy to obtain the yield in Fisher information as a function of the final parameters for both the single- and two-mode probe states. The yields for single-mode probes are
\begin{subequations}\label{eq:singlemode}
\begin{eqnarray}
	\nonumber
	J_\gamma&=&\frac{d_1^2e^{-2r}+d_2^2e^{2r}}{2\nu}+\frac{\nu^2}{\nu^2-1}\\
	\nonumber
	&&+4\left(N+\frac{1}{2}\right)^2\frac{1+\nu^2 \cosh4r}{\nu^4-1}\\
	\label{eq:singlemodegamma}
	&&-4\left(N+\frac{1}{2}\right)\frac{\nu\cosh2r}{\nu^2-1}.\\
	\label{eq:singlemodeN}
	J_N&=&4(e^\gamma-1)^2\,\frac{1+\nu^2\cosh4r}{\nu^4-1}.
\end{eqnarray}
\end{subequations}
expressed as functions of the state parameters after the action of the channel $\mc S$. In order to obtain the final values for the yield, one needs to consider the three different situations, $n=|{\bf d}_0|^2/2$, $2n+1=\nu_0$ and $2n+1=\cosh2r_0$, and substitute in Eqs.~\eqref{eq:singlemode}.

Turning to entangled probe states, we know from~\cite{monras_information_2009} that the optimal state for constraints of the form $\tr[\rho_0\,a^\dagger a]\leq n$ can only be pure. Since we are interested in evaluating the sensitivity of the different Gaussian resources, and two-mode squeezing is the only genuinely entangling resource, we will restrict ourselves to two-mode squeezed vacuum states, that we will denote by $\rho_0^\star$:
\begin{equation}
	\rho_0^\star=S_\textrm{2m}(r_0)\ket0\bra0 S^\dagger_\textrm{2m}(r_0) \, ,
\end{equation}
where $\ket0\bra0$ denotes the two-mode vacuum. $S_\textrm{2m}(r)=\exp{\frac{1}{2}(r a^\dagger b^\dagger-r^* a b)}$ denotes the two-mode squeezing operator, where $b$ is the ancillary mode. Notice that we consistently denote parameters in the probe state with the subscript $0$.
The covariance matrix for the two-mode squeezed state is
\begin{eqnarray}\label{eq:CM.tm}
	 \Sigma_0^\star&=&\frac{1}{2}\small\left(\begin{array}{cccc}\cosh2r & 0 & -\sinh2r & 0 \\0 & \cosh2r & 0 & \sinh2r \\-\sinh2r & 0 & \cosh2r & 0 \\0 & \sinh2r & 0 & \cosh2r\end{array}\right).
\end{eqnarray}
whereas the energy reads
\begin{equation}
	n=\frac{1}{2}(\cosh2r_0-1)
\end{equation}
for the two-mode entangled state. We will consistently use $n$ to compare the performances of the different classes of states.

\section{Estimating loss $\gamma$}\label{sec:gamma}
We now proceed to analyze the problem of estimating $\gamma$ alone, under the assumption that the mean photon number $N$ is known. This problem has been partially addressed in the literature \cite{venzl_quantum_2007, monras_optimal_2007, adesso_optimal_2009} with different degrees of generality. In previous studies, emphasis is placed in zero temperature channels ($N=0$). In \cite{venzl_quantum_2007} several distinct probe states are considered, always with fixed tomographic measurements $X$ and $P$, whereas~\cite{monras_optimal_2007, adesso_optimal_2009} focus their attention in the optimal probe states, respectively within the single-mode Gaussian states and non-Gaussian states. Moreover, while~\cite{venzl_quantum_2007} considers the use of entangled probes, no consideration is made about the optimality of the measurement scheme. On the other hand, References \cite{monras_optimal_2007, adesso_optimal_2009} consider optimality of both measurement and single-mode probe states, but they do not consider the use of entangled probes. In this section we will combine both approaches, namely, considering different Gaussian resources, including entanglement, while still using the powerful tools of quantum estimation theory in order to take into account the corresponding optimal measurement for each probe state.

The yield for two-mode squeezed probe states as a function of the final state parameters is a highly involved expression which provides no physical insight. Remarkably, plugging in the dependence of the final parameters as functions of the initial mean-photon number and the channel parameters provides manageable expressions, reported in Appendix~\ref{app:exact} along with the yields for single-mode states.
Notice that in both cases, two-mode squeezing is taken at fixed phase. This does not affect the generality of the analysis since single- and two-mode squeezing along different quadratures can always be taken to the standard forms~\eqref{eq:CM.sm} and \eqref{eq:CM.tm} by means of single-mode phase shifts. The phase insensitivity of the considered channels guarantees that this will not affect the yield in Fisher information.

We thus proceed to compare the different resources by writing the output parameters as functions of the initial ones~\cite{serafini_quantifying_2005}, which in turn are functions of the available energy. The resulting general expressions are exceedingly complex and provide no particular physical insight. We will, instead, explore specific parameter regimes of physical interest.
In order to present the forthcoming results in a manageable form, we will use, when convenient, the following definitions,
\begin{subequations}
\begin{eqnarray}
	x&=&n(n+1),\\
	y&=&N(N+1),\\
	z&=&e^\gamma-1.
\end{eqnarray}
\end{subequations}
\subsection{Zero-temperature baths, $N=0$}

\begin{figure}[t]
	\includegraphics[width=.45 \textwidth]{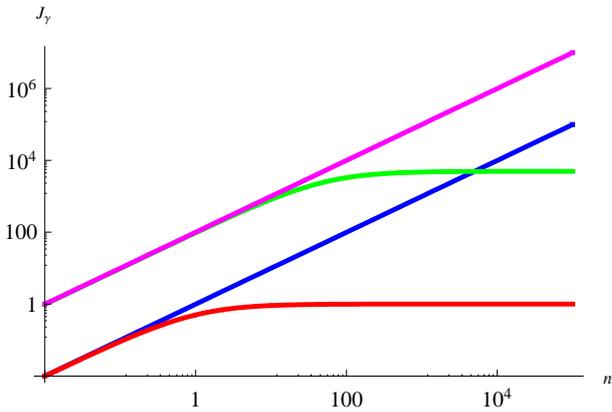}~~~~~~~~
	\caption{\label{fig:zeroN}[Color online] Log-Log plot of the yield for the $\gamma$ parameter using thermal states (red), coherent states (blue), single mode squeezed (green) and two-mode squeezed states (magenta) at different energy regimes, for $N=0$ and $\gamma=0.01$. The saturation of the performance for thermal and squeezed states is clear. On the other hand, the linear dependency of coherent and two-mode entangled states is readily visible. This is in accordance with Eqs.~\eqref{eq:zeroN}.}
\end{figure}

\begin{figure*}[t]
	\includegraphics[width=.475 \textwidth]{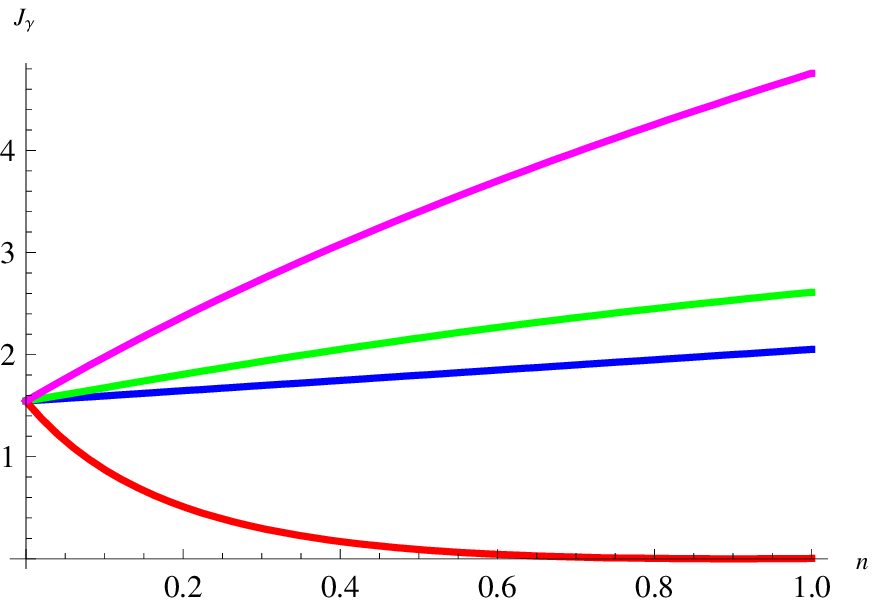}~~
	\includegraphics[width=.475 \textwidth]{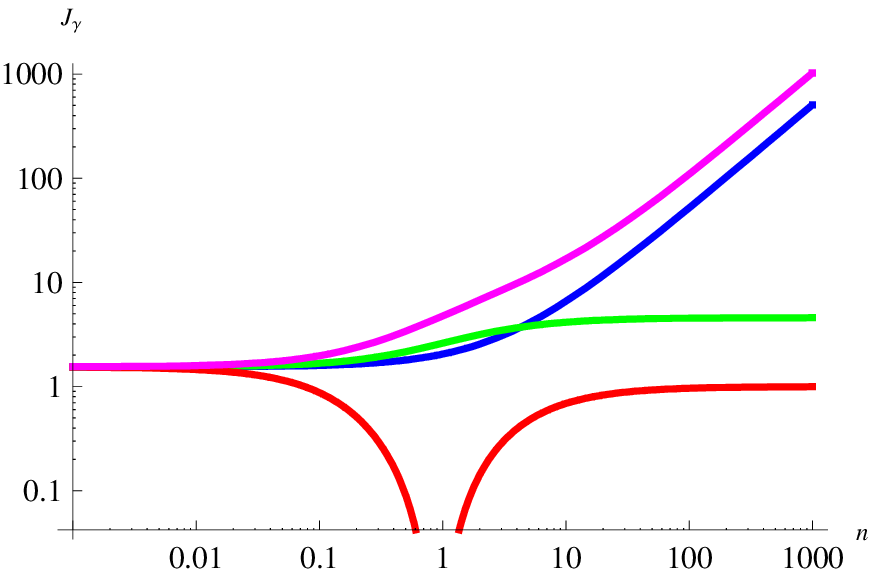}
	\caption{\label{fig:fisher_gamma}[Color online] Yield for the $\gamma$ parameter using thermal states (red), coherent states (blue), single mode squeezed (green) and two-mode squeezed states (magenta) at different energy regimes, for $N=0.9$ and $\gamma=0.3$. [Right] Log-log plot of the yield for $n$ values greater than 1. The linear behavior of coherent and two-mode squeezed states is readily apparent, whereas saturation of the yield occurs both for thermal and single-mode squeezed states. At this parameter values we have $(e^\gamma-1)(2N+1)\simeq0.98$, thus $J_\textrm{2-m}/J_\textrm{coh.}\simeq 2.0$ for $n\gg1$.~[Left] Detail of the yield at low $n$ values in linear scale. Different slopes corresponding to the different values of $J^{(1)}$ are apparent.}
\end{figure*}

As a first approach and in order to put our results in context with previous related studies \cite{venzl_quantum_2007, monras_optimal_2007, adesso_optimal_2009} we analyze the limiting case of baths at zero temperature. Zero-temperature baths are the most commonly encountered in quantum optics, and to which most of the existing literature is dedicated. Under the assumption that $N=0$, the general expressions obtained in Appendix~\ref{app:exact} reduce to
\begin{subequations}\label{eq:zeroN}
\begin{eqnarray}
	J_\text{coh.}&=& \frac{n}{z+1},\\
	J_\text{th.}&=&\frac{n}{z+1+n},\\
	J_\text{sq.}&=&	\frac{n}{z}\cdot\frac{1+z^2}{1+z(z+2(n+1))},\\
	J_{\textrm{2-m}}&=&\frac{n}{z}
\end{eqnarray}
\end{subequations}
It is easy to verify that $J_{\textrm{2-m}}$ is the largest of all these quantities. The relations $J_{\textrm{2-m}}\geq J_\text{coh.}\geq J_\text{th.}$ are obvious. On the other hand the relation $J_{\textrm{2-m}}\geq J_{\textrm{sq.}}$ only requires to observe that $(1+z^2)(1+z(z+2(n+1)))^{-1}\leq 1$.

A characteristic feature that we will encounter later on with greater generality is the fact that both thermal and single-mode squeezed states saturate their performance when $n$ is large [See Fig.~\ref{fig:zeroN}]. On the other hand, the performance of coherent and two-mode squeezed states grows linearly with $n$. This clearly leaves the latter as the two candidates for optimality when $n$ is large. A relevant question is when does the performance of one become \emph{much} larger than the other's? As will be seen, two-mode squeezed vacuum states perform always better than coherent ones. On the other hand, it is easy to see that the increase in performance of two-mode squeezed states is most relevant when $z\ll z+1$ ($\gamma\ll1$), since then $J_{\textrm{2-m}}\gg J_{\textrm{coh.}}$. This increase in performance is independent of the amount of energy in the probe state. This can be observed also from Fig.~\ref{fig:zeroN}.
%
%

\subsection{Low energy regime}
We now turn to the most general situation where the thermal bath has nonzero temperature, \emph{i.e.} photons can \emph{leak into} the quantum system additionally to \emph{leaking out} from it. We consider the two interesting parameter regimes with practical relevance, namely, that of low-energy probes and that of high energy probes. The former is best suited for situations where the properties of the channel (bath) under inspection are sensitive to the effect of intrusive probing. It is worth stressing that in some cases, the gain by using a small amount of energy may not provide a substantial gain w.r.t. the performance of the vacuum. On the other hand, there are situations where the choice of the probe state critically determines the attainable accuracy. We wish to identify those situations.

In the low energy regime ($n\ll1$) we perform the Taylor expansion
\begin{equation}
	J_\gamma=J^{(0)}+J^{(1)} n+o(n^2),
\end{equation}
where obviously $J^{(0)}$ is independent of the kind of state being considered, and corresponds to the performance of the vacuum. The common leading constant is thus given by
\begin{equation}
	J^{(0)}=\frac{N/z}{1+z(N+1)}.
\end{equation}
Corrections of order $n$ contribute with coefficients
\begin{subequations}
\begin{eqnarray}
	J^{(1)}_\text{coh.}&=&\frac{1}{1+z(1+2N)},\\
	J^{(1)}_\text{th.}&=&-\frac{(z+1) (1+2z(N+1))}{z^2 (1+z(N+1))^2},\\
\nonumber
	J^{(1)}_\text{sq.}&=&	\frac{2N+1}{z}-\frac{z(1+N)^2(1+2N)}{(1+z(N+1))^2}
	\\
	&&+\frac{2(1+2N)^2}{(1+z)^2+2Nz(1+z(N+1))}\\
	J^{(1)}_{\textrm{2-m}}&=& \frac{(z+1)^2+N (z (z+2)+2)}{z (1+z(N+1))^2}
	\end{eqnarray}
	\end{subequations}
where we have defined $y=N(N+1)$. A few comments are in order. First of all, notice from Eq.~\eqref{eq:exact.gamma.coherent} that the yield for coherent states is a polynomial of first degree in $n$. Therefore, $J^{(0)}+J_\textrm{coh.}^{(1)}n$ gives the exact expression. On the other hand, the thermal correction $J^{(1)}_\textrm{th.}\leq 0$ is always negative, which implies that weak thermal fields perform worse than the vacuum [See Fig.~\ref{fig:fisher_gamma}~(Left)].

The first question to ask is when do coherent states provide any significant improvement over the vacuum. For this we consider the condition $J^{(0)}\ll J_\textrm{coh.}^{(1)}n$. This reduces to
\begin{equation}\label{eq:coherent.cond}
	\frac{N(z(2N+1)+1)}{z(z(N+1)+1)}\ll n.
\end{equation}

On the other hand, the same condition for two-mode squeezed probe states $J^{(0)}\ll J^{(1)}_\textrm{2-m}n$, reduces to
\begin{equation}\label{eq:2m.cond}
	\frac{N(z (N+1)+1)}{(N+1)(z+1)^2+2N}\ll n.
\end{equation}
Notice that for moderate values of $N$ and large values of $\gamma$ ($z\gg1$) Eqs.~\eqref{eq:coherent.cond} and \eqref{eq:2m.cond} reduce to $N (2N+1)/(z(N+1))\ll n$ and $N/z\ll n$ respectively, which shows that the regime required for Eq.~\eqref{eq:2m.cond} is entered earlier than that of Eq.~\eqref{eq:coherent.cond} for increasing $z$.

In the limit of small losses ($z\ll1$), we can expand Eqs.~\eqref{eq:coherent.cond} and  \eqref{eq:2m.cond} to zeroth order in $z$ to obtain, for $J^{(0)}\ll J_\textrm{coh.}^{(1)}n$
\begin{equation}\label{eq:coherent.cond2}
	\frac{N}{z}+N^2\ll n
\end{equation}
whereas for $J^{(0)}\ll J^{(1)}_\textrm{2-m}n$  we get
\begin{equation}
	\frac{N}{1+2N}\ll n.
\end{equation}
The condition for $J^{(0)}\ll J^{(1)}_\textrm{2-m}n$ thus reduces to $N\ll n/(1-2n)\simeq n$. Notice that this is not a sufficient condition to achieve a significant improvement using coherent probes [Eq.~\eqref{eq:coherent.cond2}].

On the other hand, one can see that $\gamma\ll1$ (that is, $z\ll1$) with moderate $N$ ($z(1+2N)\ll1$) implies $J_\textrm{coh.}^{(1)}\ll J_\textrm{2-m}^{(1)}$. This means that at moderate temperatures, and with low energy in the probe state, two-mode squeezed states significantly outperform coherent states in the regime of small losses.

%
%
%
\begin{figure*}[t]
	\includegraphics[width=.475 \textwidth]{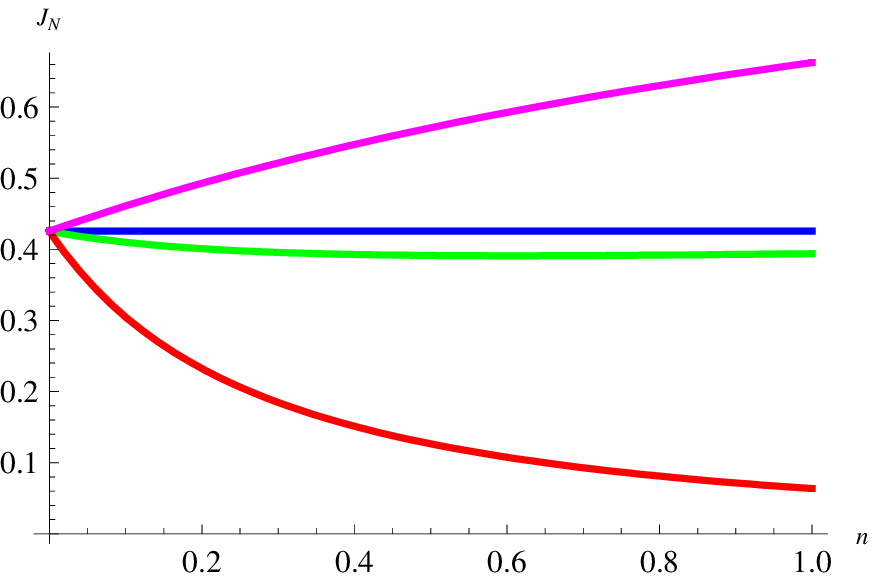}~~
	\includegraphics[width=.475 \textwidth]{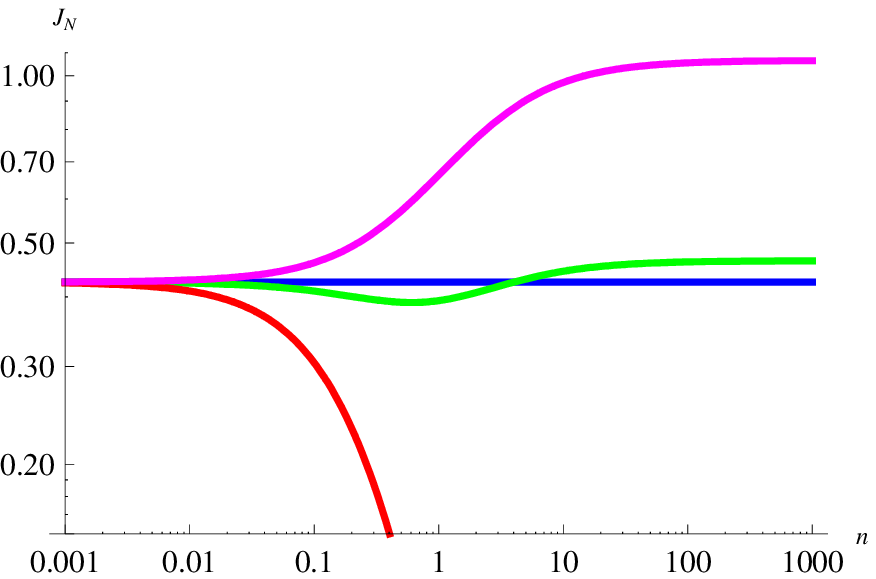}
	\caption{\label{fig:fisher_N}[Color online] Yield for the $N$ parameter using thermal states (red), coherent states (blue), single mode squeezed (green) and two-mode squeezed states (magenta) at different energy regimes, for $N=0.9$ and $\gamma=0.3$.
	[Above] Detail of the yield at low $n$ values.  The $X$ parameter is negative, which implies that single-mode squeezing is detrimental (at small $n$ values).
	[Below] Behavior at high $n$ values.
}
\end{figure*}
\subsection{High energy regime}
High energy probes are of interest when the channel being probed is not as delicate, or needs not be preserved. A natural instance of this situation is in probing the photon-photon scattering predicted by QED and non-standard models of elementary particles.
The high energy regime has a substantially different behavior. Expanding the relevant yield functions from Appendix~\ref{app:exact} in inverse powers of $n$ we obtain series of the form
\begin{equation}
	J_\gamma=J^{(-1)}n+J^{(0)}+o(1/n)
\end{equation}
In some cases $J^{(-1)}$ will vanish, rendering the corresponding class of states useless compared to those for which $J^{(-1)}$ does not vanish. Explicitly, we have
\begin{subequations}
\begin{eqnarray}
	J_\text{coh.}&=&\frac{n}{1+z(2 N+1)}+O(1),\\
	J_\text{th.}&=&1+o\left(\frac{1}{n}\right),\\
	J_\text{sq.}&=&\frac{1}{2} \left(1+\frac{1}{z^2}\right)+o\left(\frac{1}{n}\right)\\
	J_{\textrm{2-m}}&=&\frac{n}{z(2 N+1)}+O(1)
\end{eqnarray}
\end{subequations}
The first relevant fact to notice is that, contrary to the low energy regime, different Gaussian resources perform differently in the limit of large energy. This is hardly a surprise. We observe that the asymptotic performance is bounded for thermal and single-mode squeezed states. This is in contrast to the fact that single-mode squeezed states have proven highly efficient for other precision measurements such as optical phase~\cite{caves_quantum-mechanical_1981, klauder_squeezed_states, monras_optimal_2006} and magnetometry~\cite{molmer_estimation_2004}, among others. On the other hand, coherent and two-mode squeezed states provide an unbounded yield, as is manifest by the nonvanishing $J^{(-1)}$ terms, giving a linear growth in $J_\gamma$ with increasing $n$. This makes the constant zeroth order correction irrelevant. A very important difference between coherent states and two-mode squeezed states becomes readily apparent. While for coherent states, the \emph{rate} of growth of $J^{(-1)}$ is bounded, for two-mode squeezed states it is not. In particular, the difference between the two yields becomes most significant when $z(2 N+1)\ll1$ and negligible when $z(2 N+1)\gg1$. This is certainly relevant for detecting very small damping parameters, for which the inverse dependence in $z$ may even be sufficient to overcome the practical limitations of achieving very high $n$ values.

\section{Estimating temperature}\label{sec:N}
Quantum thermometry has become a subject of high physical relevance with the advent of ultracold atomic gases~\cite{hofferberth_probing_2008,gottlieb_quantum_2009,manz_two-point_2010}. At low temperatures, new methods need to be envisaged to determine the magnitude of thermal fluctuations in atomic clouds. In this section we analyze the quantum-limited precision bounds to the estimation of temperature (mean photon number $N$) in a bosonic thermal bath, coupled to a probe system prepared in a Gaussian state, with the coupling strength not necessarily large, \emph{i.e.}, far from thermalization.

A straightforward method to measure temperature is to let the probe system coupled to the bath to thermalize. Quadrature measurements then provide an estimator of the mean photon number, providing in turn an estimator of the bath temperature. This approach has several drawbacks. Most importantly, it requires in general a large coupling constant $\gamma\gg1$, in order to reach the steady state. However, there may be situations where the coupling constant cannot be chosen at will. On the other hand, this does not necessarily provide the optimal estimation accuracy. As we will see, two-mode squeezed states can outperform the sensitivity of the vacuum state or other classes of Gaussian states, for any value of the parameters.

Contrary to the effect of the decay parameter $\gamma$, that affects both first and second moments, the temperature in the bath only affects the second moments. First moments evolve independently of the bath temperature~\cite{serafini_quantifying_2005}. This has immediate consequences for the sensitivity of coherent states, which will perform equivalently to the vacuum state. Thus, we will not consider coherent probe states in this section. We will follow the same approach taken in the previous section, by addressing different energy regimes in the probe states.

\subsection{Low energy regime}

\begin{figure*}[t]
	\includegraphics[width=.45 \textwidth]{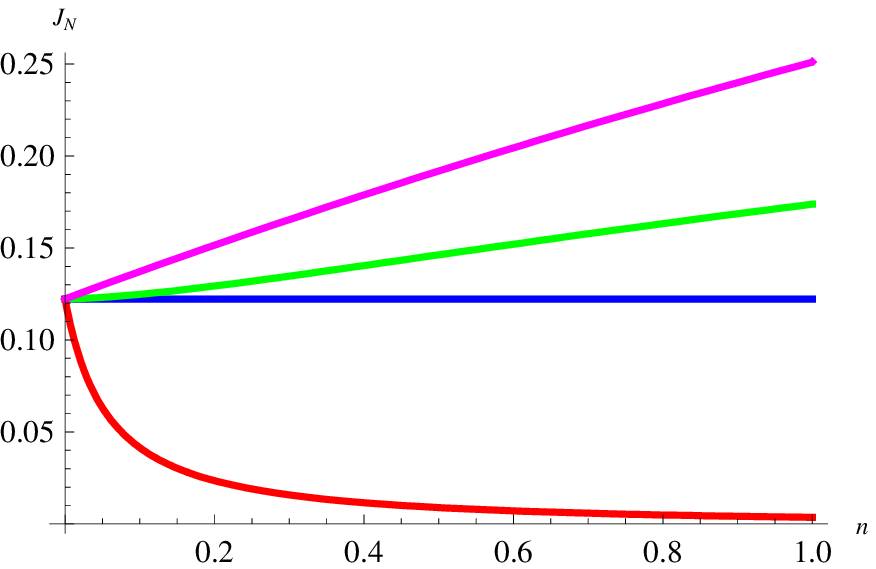}~~~~~~~~
	\includegraphics[width=.45 \textwidth]{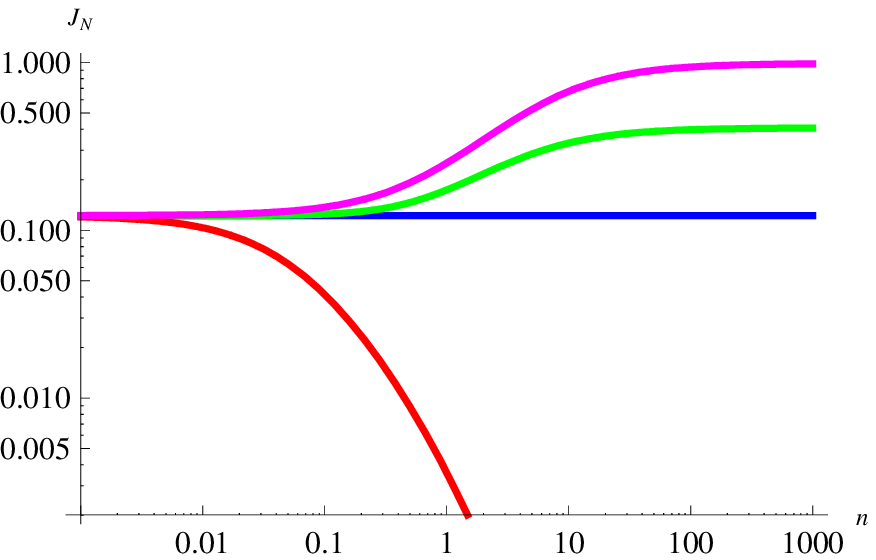}
	\caption{\label{fig:fisher_N2}[Color online] Yield for the $N$ parameter using thermal states (red), coherent states (blue), single mode squeezed (green) and two-mode squeezed states (magenta) at different energy regimes, for $N=0.7$ and $\gamma=0.08$.
	[Left] Detail of the yield at low $n$ values.  The $X$ parameter is positive, which can be see by the positive slope of the single-mode squeezing ($n\ll1$).
	[Right] Behavior at high $n$ values.
}
\end{figure*}

As in the case for $J_\gamma$, $J_N$ will be limited to a constant value corresponding to the sensitivity of the vacuum state $J^{(0)}$ at small values of $n$. However, since first moments are unaffected by the temperature of the bath, the yield of coherent states will equal that of the vacuum. As in the previous section, we will expand $J_N$ in powers of $n$, in order to obtain and analyze the low-energy yield for each class of states,
\begin{equation}
	J_N=J^{(0)}+J^{(1)}n+o(n^2)
\end{equation}
Taking the expressions from Appendix~\ref{app:exact}  we get
\begin{equation}
	J^{(0)}=J_\textrm{coh.}=\frac{z (z+1)^2}{N (1+z(N+1))}
\end{equation}
to which corrections of order $n$ contribute with factors
\begin{subequations}
\begin{eqnarray}
	J^{(1)}_\text{th.}&=&-\frac{(z+1)^2 (z(2 N+1)+1)}{N^2 (z(N+1)+1)^2},\\
\nonumber
	J^{(1)}_\text{sq.}&=&
	\frac{32 z (z+1)^2 \left(2(\xi-1)-z \left(4 z \xi ^3+(z+2) \xi +1\right)\right)}
	{\left(4 z \xi ^2+4 \xi -z-2\right)^2\left(z \left(4 z \xi ^2+4 \xi +z+2\right)+2\right)}\\
	&&\\
	J^{(1)}_{\textrm{2-m}}&=&\frac{(2 N+1) z (z+1)^2}{N\left(N+1\right) (z(N+1)+1)^2},
\end{eqnarray}
\end{subequations}
where we have defined $\xi=N+1/2$.

It is immediate to observe that $J^{(1)}_\text{th.}\leq0$ which implies that, similarly to the situation for $J_\gamma$, small thermal fluctuations in the probe state can only be detrimental. On the other hand, the small $n$ correction for single-mode squeezed states has no unambiguously defined sign,  the latter being positive only when $X=2(\xi-1)-z \left(4 z \xi ^3+(z+2) \xi +1\right)>0$. Solving the inequality $X>0$ for $z$ we obtain
\begin{equation}
	z<\frac{(2 \xi-1) \sqrt{8 \xi ^2+1} -2 \xi-1
	}{2\xi(4 \xi ^2+1) }.
\end{equation}
The right hand side is positive only when $\xi>1$, which means that only for $N>1/2$ it is possible to have a gain in yield by using single-mode squeezed
probes. We thus conclude that for $0<N<1/2$ the best single-mode probe (at low energies, $n\ll1$) is the vacuum state. However, if the bath temperature is sufficiently high [$N>1/2$], it is possible to improve the sensitivity of the vacuum by single-mode squeezing.

We now turn to analyze the yield of two-mode squeezed states. As can be readily seen in Figs.~\ref{fig:fisher_N} and \ref{fig:fisher_N2}, the behavior is rather simple. The slope at small values of $n$ is always positive and, moreover, as follows from the relations in Appendix~\ref{app:exact}, always larger than that of single-mode squeezed states. Imposing that $J_\textrm{2-m}^{(1)}n\gg J^{(0)}$ yields $n(2N+1)\gg z(N+1)^2+N+1$. Since the right hand side is greater than unity, we have $n(2N+1)\gg 1$ and, given that $n$ is small, it follows that $N$ has to be large. Therefore we can reduce the condition to $z\ll(2n-1)/N$, which is never satisfied for small $n$. We thus conclude that low energy entangled probes always outperform single-mode ones, for any regime of the parameters, although the improvement cannot always be of significant magnitude ($J_\textrm{2-m}^{(1)}n\gg J^{(0)}$ cannot be achieved for small $n$ values). Most remarkably, two-mode squeezed states are the only class of Gaussian states that perform better than the vacuum for any values of the parameters.

\subsection{High energy regime}
The high energy limit for $J_N$ is quite uninteresting. The reason being that all yields saturate, and no significant improvement can be achieved using two-mode squeezed probes as compared to the use of coherent states. The limiting expressions read, to order $1/n$
\begin{eqnarray}
	J_\text{th.}&=&\frac{z^2(1+z)^2}{n^2}\simeq 0,\\
	J_\text{sq.}&=&\frac{2 (1+z)^2}{(2 N+1)^2}-\frac{(1+z)^2}{(2 N+1)^3zn}\\
	J_{\textrm{2-m}}&=&\frac{(1+z)^2}{N(N+1)}-\frac{(1+z)^2}{N\left(2 N^2+3 N+1\right)
   zn}
\end{eqnarray}
We have presented the second order term for $J_\textrm{th.}$ because it is the leading one. The yield, however, tends to vanish, as can be expected by observing that in the limit of a highly energetic probe, the thermal fluctuations in the probe are infinitely larger than those induced by the bath, and therefore no inference about the latter can be obtained. Single- and two-mode squeezed states have nonzero limiting yields, but, as anticipated, the yield saturates for highly energetic probes. It is easy to check that in the limit $n\rightarrow\infty$ we have $J_{\textrm{2-m}}\geq J_\text{sq.}$ but no order can be established between the vacuum state and single-mode squeezed states. A crossover can occur between the performances for the vacuum and single-mode squeezed states depending on the values of $\gamma$ and $N$.

Concentrating on the asymptotic yield, a natural question is to understand in what situations does two-mode squeezing perform much better than the vacuum state. This is simple to answer by observing that
\begin{equation}
	\frac{\lim_{n\rightarrow\infty}J_{\textrm{2-m}}}{J_\textrm{coh.}}=1+\frac{1}{(N+1)z}
\end{equation}
so that if $(N+1)z\ll1$ (in the high $n$ limit) then $J_{\textrm{2-m}}\gg J_\textrm{coh.}$. This condition is likely to occur in several practical situations, for channels close to ideal, \emph{i.e.}, when the decay rate $\gamma$ is very low and the temperature is low or moderate. Notice that this is not the first time that we encounter this condition for a significant improvement of squeezed states over coherent ones. All results indicate that whenever there is an improvement of two-mode squeezed states over the vacuum, it is always in the limit of small coupling to the bath ($\gamma\ll1$, i.e., a close to ideal channel).

\section{Entanglement and the performance of Non-Gaussian states}\label{sec:nonG}
\begin{figure*}[t]
\includegraphics[width=.46\textwidth]{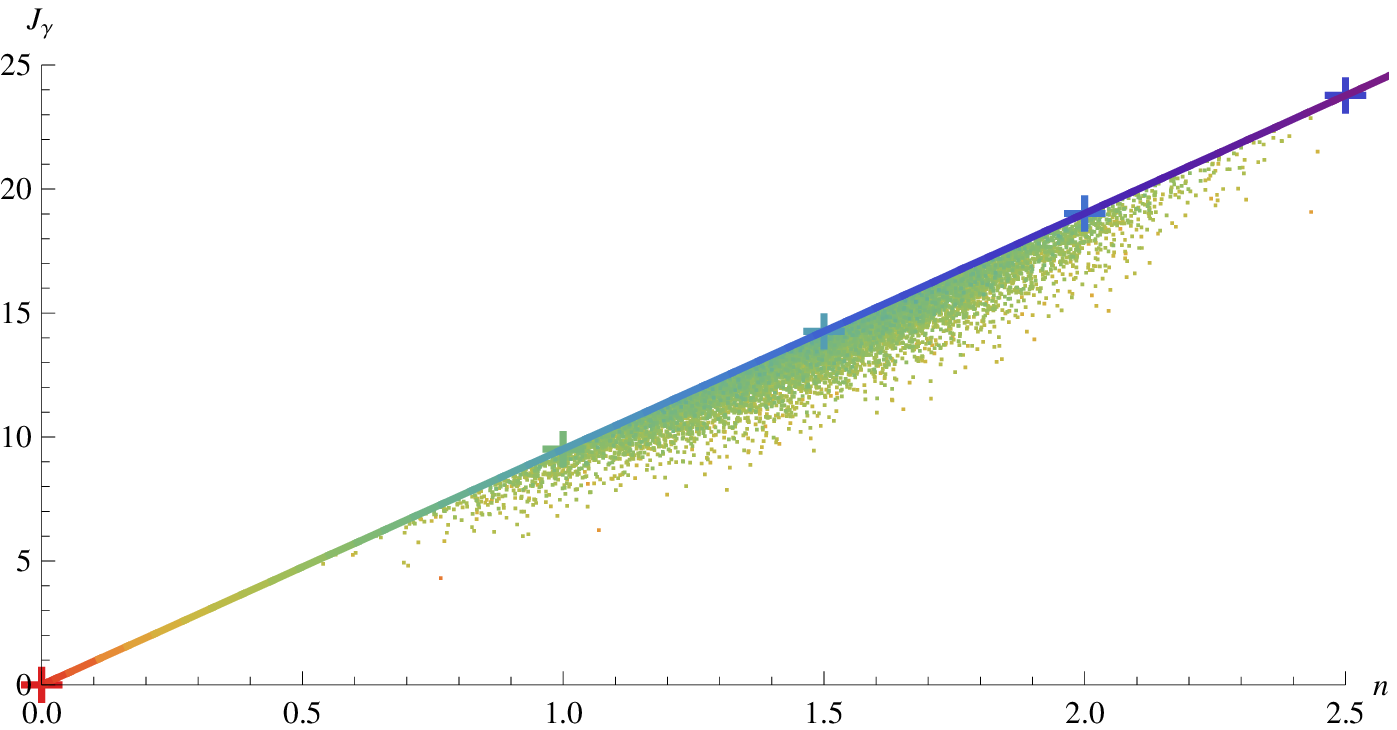}~~~~
\includegraphics[width=.5\textwidth]{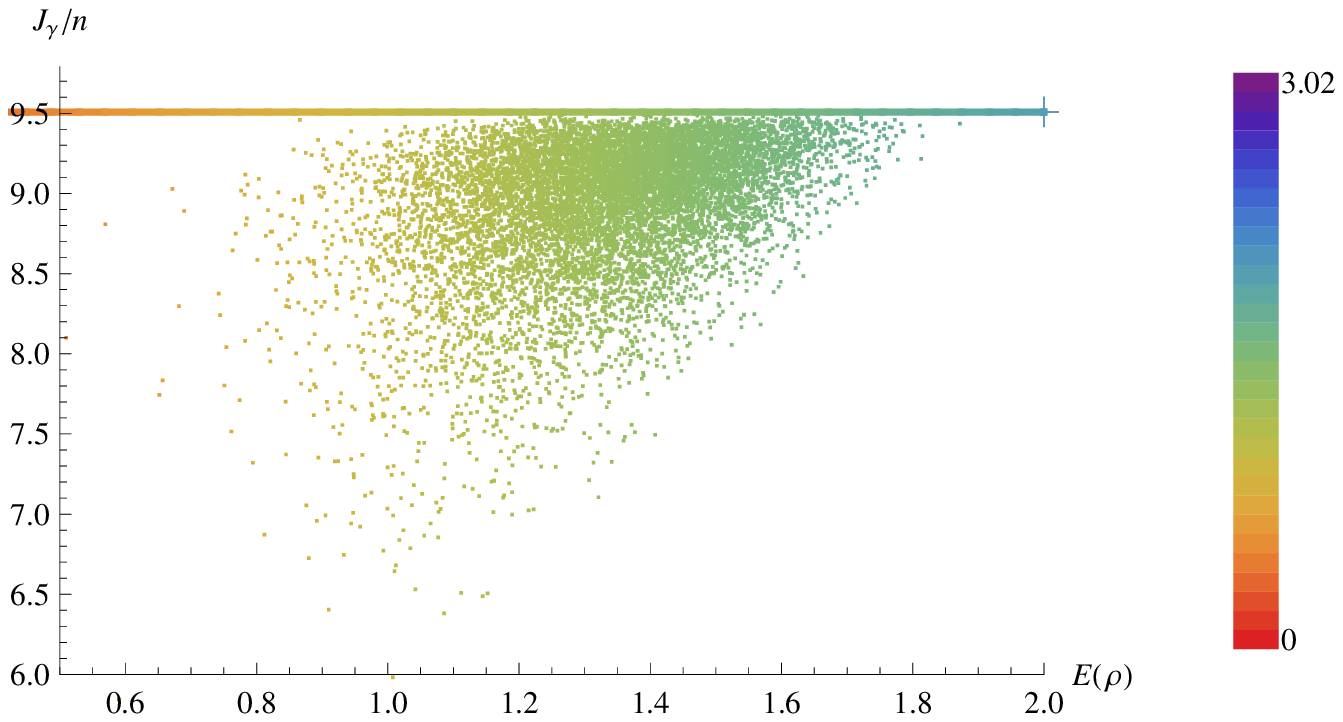}
\caption{\label{fig:random} [Color online] Scatter plot of 4000 random probe states with at most 3 photons in each mode (dots), maximally entangled states at dimension cutoff $3\leq d\leq 6$, $\ket\psi=\frac{1}{\sqrt d}\sum_k\ket{k}\ket{k}$ (crosses), and two-mode squeezed vacuum states (solid line). \emph{Left}: Yield against mean photon number in the channel mode $a$ as a reference. The dashed line interpolates the behavior of maximally entangled states. Color code corresponds to the entropy of entanglement in the probe state $E(\rho)=-\tr\rho_a\log\rho_a$, $\rho_a=\tr_b \rho$. \emph{Right}: Ratio between the yield and the mean photon number against the entropy of entanglement. Observe that several highly efficient probes (those with high ratio $J_\gamma/n$) are relatively unentangled as compared to the maximally entangled ones with similar performance. Clearly, the squeezed vacuum state is much more entangled than randomly sampled states with similar yield.}
\end{figure*}

So far we have seen that, in order to optimally estimate the channel parameters in Eq.~\eqref{eq:channel}, two-mode squeezing is the most effective resource within the arena of Gaussian inputs. This fact suggests two immediate questions. \emph{1)} Is there a direct relation between entanglement and the performance in channel estimation? And \emph{2)} Are there non-Gaussian states outperforming the squeezed vacuum for the same energy supply? Questions similar to \emph{1)} have arisen previously in the literature, both in the finite dimensional case~\cite{fujiwara_quantum_2003,fujiwara_estimation_2004, boixo_operational_2008} and in the continuous variable one (albeit in somewhat different setups~\cite{tan_quantum_2008}), considering different kinds of channels. Question \emph{2)} has been also addressed in the context of quantum channel estimation~\cite{adesso_optimal_2009}, and regarding the performance in other quantum information tasks, especially continuous-variable quantum teleportation~\cite{dellanno_continuous-variable_2007}.

With the available techniques it is difficult to give a precise \emph{quantitative} answer, since numerical techniques are not well-suited to deal with infinite dimensional systems, and analytic control methods are not yet well developed beyond the Gaussian regime. The difficulty with numerical methods resides in the fact that Hilbert-space truncation is rendered useless at $N\neq0$, because thermal baths immediately populate all levels in Fock space, thus rendering a direct numerical approach futile, or, in the best of all cases, a rude approximation. In order to address the question of sensitivity of non-Gaussian states for parameter estimation, techniques beyond those developed so far are needed, and it is beyond the scope of this work to pursue them.
Instead, we will provide a simple \emph{qualitative} answer by restricting the channels of interest to those at zero temperature, \emph{i.e.}, $N=0$. In this case, populated levels in Fock space only decay to lower energy levels and Hilbert-space truncation provides exact numerical results.

We have computed numerically the quantum Fisher information $J_\gamma$ at $\gamma=0.1$ for $4000$ states picked from $\mathbb C^4\otimes\mathbb C^4$ constituting the subspace with at most 3 photons in each mode, randomly distributed according to the $SU(16)$ Haar measure. In Fig.~\ref{fig:random} we report our findings. The left plot displays the values of $J_\gamma$ against the mean photon number in mode $a$, and the right plot displays the \emph{efficiency} (\emph{i.e.}, the ratio $J_\gamma/n$) against the entropy of entanglement $E(\rho)=-\tr\rho_a \log \rho_a,~\rho_a=\tr_b\rho_0$. Along with the random states we display the results for maximally entangled states within the cutoff $3\leq d\leq 6$ and the two-mode squeezed vacuum. Our findings reveal that most states have a very high performance in relation to the amount of energy they have in the channel mode. In particular, for a fixed dimension cutoff the more entangled the states are, the better they perform on average. In particular, the maximally entangled state attains the bound set by the squeezed vacuum.
It is of course interesting to ask how does the performance change when one introduces temperature in the channel. This is beyond the capabilities of numerical methods relying on dimensional cutoff, and more advanced methods would need to be envisaged.

\section{Discussion}\label{sec:discuss}
We have obtained the sharp precision bounds on estimation of $\gamma$ and $N$ for four classes of Gaussian states, namely, coherent, thermal, single-mode squeezed vacuum and two-mode squeezed vacuum. We have shown that the two-mode squeezed vacuum always outperforms any other class of Gaussian states. The improvement of two-mode squeezed vacuum states versus coherent states is most relevant when the coupling parameter $\gamma$ is weak. In particular, at zero or finite temperature, the yield $J_\gamma$ of two-mode squeezed states increases much faster with $n$ than the yield of coherent states at small values of $\gamma$. For $J_N$, comparing the yield of two-mode squeezed vacuum states versus the yield of the vacuum, we find that, for small values of $n$, no significant improvement can be obtained. The situation changes dramatically in the high energy regime, where, despite a saturation of the yield (all yields saturate to a maximum value), saturation occurs at much higher yields when $z(N+1)\ll1$. Summarizing, we have shown that two-mode squeezed vacuum always outperforms any other Gaussian resource for both estimation problems ($\gamma$ and $N$),  and we have identified the situations in which this improvement is most significant.
We have also provided numerical evidence that the squeezed vacuum state provides an upper bound to the performance of arbitrary states with dimensional cutoff. This suggests a deeper analysis, and the causes for this optimality should be investigated.

Turning to the multiparametric problem of simultaneously estimating both parameters of the channel, we have shown that the optimal state (two-mode squeezed vacuum) provides optimal sensitivity for both parameters, and that in the many-copy limit ($k\rightarrow\infty$) a collective measurement exists which saturates both bounds simultaneously, as a consequence of Eq.~\eqref{EQ:COMMUTE}. These results imply that the problems of estimating $\gamma$ and $N$ enjoy a strong form of compatibility. Not only the optimal probe state is common to both problems, but also the corresponding optimal measurements commute in the asymptotic limit. Therefore an overall protocol optimizing both tasks can be envisaged (in the asymptotic $k\rightarrow\infty$ limit). It is a relevant question to determine whether lifting the restriction to Gaussian states can provide an increased performance, and whether such kind of compatibility remains true in the more general setting
involving arbitrary probe states.

~\\*
\noindent\textbf{Acknowledgements.} The authors are thankful to Dr. M. Gu\c{t}\u{a} for very helpful discussions. We acknowledge financial support from the European Commission of the
European Union under the FP7 STREP Project HIP (Hybrid Information Processing),
Grant Agreement n. 221889.

\appendix
\section{Proof of Eq.~\eqref{EQ:COMMUTE} for two-mode squeezed vacuum states}\label{sec:commute}
In this section we show that, for two-mode squeezed vacuum probe states, Eq.~\eqref{EQ:COMMUTE} holds. Since we are probing channel $\mc S^\star$ with bipartite states we will be working with two bosonic modes.
Start by introducing some notation. Let $\langle X\rangle=\tr[\rho X]$ be the expectation value of any operator, evaluated with the state output from the channel $\rho=\mc S^\star\rho_0$. $R^i=(Q_1,P_1,Q_2,P_2)$ are the canonical operators with $[R^i,R^i]=i\Omega^{ij}$ and $\Omega$ is the 2-mode symplectic matrix, $\Omega=
\omega\oplus\omega$, with
\begin{equation}
	\omega=\left(\begin{array}{cc}0 & 1 \\-1 & 0\end{array}\right).
\end{equation}
It is convenient to work with the centered canonical operators $\tilde R^i=R^i-\langle R^i\rangle$ with which we define the covariance matrix of the output state $\Sigma^{ij}=\langle \tilde R^i\circ \tilde R^j\rangle$. Here we have introduced the symmetric product $A\circ B=(AB+BA)/2$. We will be using Einstein's summation convention and tensors of the form $A^i_{\,k}\,B^j_{\,l}$ will be written as $[A\otimes B]^{ij}_{~kl}$. Analogously $A^i_{~k}v^j$ will be written $[A\otimes v]^{ij}_{~k}\equiv[v\otimes A]^{ji}_{~k}$, if no confusion arises. Finally, we construct a covariance matrix with covariant-contravariant transformation rules, namely $\tilde\Sigma^i_{~k}=\langle\tilde R^i\circ \tilde R^j\rangle \Omega_{jk}$ so that $\tilde\Sigma=\Sigma\Omega$.

The covariance matrix of the input state will be denoted $\Sigma_0$,
\begin{equation}\label{eq:SVS}
	\Sigma_0=\frac{1}{2}\left(\begin{array}{cccc}\cosh r & 0 & \sinh r & 0 \\0 & \cosh r & 0 & -\sinh r \\\sinh r & 0 & \cosh r & 0 \\0 & -\sinh r & 0 & \cosh r\end{array}\right)
\end{equation}
and observe that the final covariance matrix is in standard form~\cite{duan_inseparability_2000},
\begin{equation}\label{eq:stdform}
	\Sigma=\left(\begin{array}{cc}a\openone & c \ZZ \\c
\ZZ & b\openone
\end{array}\right)
\end{equation}
where $\ZZ=\diag(1,-1)$, $a=e^{-\gamma}\cosh r+(1-e^{-\gamma})(N+1/2)$, $b=\cosh r$ and $c=a^{-\gamma/2}\sinh r$. Next, define the superoperator $\mc D_\mu$ as $\mc D_\mu \mc S^\star = \partial_\mu\mc S^\star$ and notice that, for both parameters $\gamma$ and $N$ it holds that~\cite{monras_information_2009}
\begin{equation}
	\mc D_\mu \rho=\alpha_{\mu,ij}(R^i \rho R^j- (R^jR^i)\circ\rho)
\end{equation}
where the $\alpha$ matrices have the diagonal block structure
\begin{eqnarray}
	\alpha_\gamma=\left(
	\begin{array}{cccc}
		N+{1/2} & -{i/2} &&\\
		{i/2} & N+1/2&&\\
		&&0& \\
		&&&0 \end{array}\right),\\
	\alpha_N=(e^\gamma-1)\left(
	\begin{array}{cccc}
	1 & && \\
	& 1&&\\
	&&0&\\
	&&&0
	\end{array}\right).
\end{eqnarray}
With this, the SLD's read~\cite{monras_information_2009}
\begin{equation}
	\Lambda_\mu=\alpha_{\mu,ij}\mc L^{ij}\\
\end{equation}
where
\begin{equation}\label{eq:SLD2}
	\mc L^{ij}=
		\big[L^{(0)}\big]^{ij}_{~kl}\Omega^{kl}+\big[L^{(1)}\big]^{ij}_{~k}\tilde R^{k}
		+\big[L^{(2)}\big]^{ij}_{~kl}
		\tilde R^{k}\circ \tilde R^{l},
\end{equation}
and
\begin{eqnarray}
	 L^{(0)}&=&i\,D^{-1}\Big(\tilde\Sigma\otimes\tilde\Sigma-\frac{i}{4}(\tilde\Sigma\otimes\openone-\openone\otimes\tilde\Sigma)\Big),\\
 \label{eq:L}
	L^{(1)}&=&\,2i\big(\tilde\Sigma^{-1}\otimes\langle R\rangle-\langle R\rangle\otimes\tilde\Sigma^{-1}\big),\\
	L^{(2)}&=&\frac{1}{2}D^{-1}\big(
			\openone\otimes\openone-i(\tilde\Sigma\otimes\openone-\openone\otimes\tilde\Sigma)
		\big),
\end{eqnarray}
and we have defined
\begin{equation}
	\label{eq:Ddef}
	D=\tilde\Sigma\otimes\tilde\Sigma-\frac{1}{4}\openone\otimes\openone.
\end{equation}
Also, observe that both $\alpha_\mu$ matrices share the same structure $\alpha_\mu=\kappa_\mu P +  i \iota_\mu Q$ where
\begin{equation}
P=\left(
	\begin{array}{cc}\openone & 0 \\
0&0 \end{array}\right),
\quad
 Q=\left(\begin{array}{cc}\omega &0 \\
  0 & 0\end{array}\right).
\end{equation}

In order to evaluate Eq.~\eqref{EQ:COMMUTE}, observe that some terms in Eq.~\eqref{eq:SLD2} immediately drop when taking the commutator or the expectation value
\begin{equation}\label{eq:expr}
	\tr[\rho[\Lambda_\mu,\Lambda_\nu]]=\alpha_{\mu,\,ij}\alpha_{\nu,\,i'j'}T^{iji'j'},
\end{equation}
where
\begin{eqnarray}
	T^{iji'j'}&=&\tr[\rho [\mc L^{ij},\mc L^{i'j'}]]\\
	&=&\big[L^{(1)}\big]^{ij}_{~k}\big[L^{(1)}\big]^{i'j'}_{~k'}\Omega^{kk'}\\
	\nonumber
	&&+\big[L^{(2)}\big]^{ij}_{~kl}\big[L^{(2)}\big]^{i'j'}_{~k'l'}\langle [\tilde R^k\circ \tilde R^l,\tilde R^{k'}\circ \tilde R^{l'}]\rangle.
\end{eqnarray}
Proceed by evaluating $T^{iji'j'}$. The following relations will be used,
\begin{align}
	\tr[\rho [\tilde R^i,\tilde R^{i'}]]=&\,i\,\Omega^{ii'},\\
	\tr[\rho [\tilde R^i,\tilde R^{i'}\circ \tilde R^{j'}]]=&\,0,\\
	\nonumber
	\tr[\rho [\tilde R^i\circ\tilde R^j,\tilde R^{i'}\circ \tilde R^{j'}]]=&\,i\big(\Omega^{ii'}\Sigma^{jj'}+\Omega^{ji'}\Sigma^{ij'}\\
\label{eq:bicommutator}
	&+\Omega^{ij'}\Sigma^{ji'}+\Omega^{jj'}\Sigma^{ii'}\big).
\end{align}
Let us define the quartic momenta $K^{iji'j'}=\langle [\tilde R^k\circ \tilde R^l,\tilde R^{k'}\circ \tilde R^{l'}]\rangle$, and the superoperator (in phase-space) $\mathfrak D$  as $\DD[A]_{ij}=A_{i'j'}D^{i'j'}_{~~ij}$. Notice that $\DD$ is linear and $\DD[A^\top]=\DD[A]^\top$, thus $\DD[A]$ preserves the symmetric or antisymmetric character of $A$, and so does any function of $\DD$, such as $\DD^{-1}$. With this we define
$\PP=\mathfrak D^{-1}[P]$ and $\QQ=\mathfrak D^{-1}[Q]$, and we can write Eq.~\eqref{eq:expr} as
\begin{subequations}
\begin{align}\nonumber
\alpha_{\mu,ij}&\alpha_{\nu,i'j'}T^{iji'j'}\\
\label{eq:linearterms}
=&\alpha_{\mu,ij}\alpha_{\nu,i'j'}\big[L^{(1)}\big]^{ij}_{~k}\big[L^{(1)}\big]^{i'j'}_{~k'}\Omega^{kk'}\\
\label{eq:real}
	&(\kappa_\mu\kappa_\nu\PP_{ij}\PP_{i'j'}-\iota_\mu\iota_\nu \QQ_{ij}\QQ_{i'j'}){\mc N}^{ij}_{kl}{\mc N}^{i'j'}_{k'l'}K^{klk'l'}\\
\label{eq:imag}	
&+i(\kappa_\mu\iota_\nu-
	\iota_\mu\kappa_\nu)\PP_{ij}\QQ_{i'j'}{\mc N}^{ij}_{kl}{\mc N}^{i'j'}_{k'l'}K^{klk'l'}
\end{align}
\end{subequations}
Concentrating on line \eqref{eq:linearterms}, it is easy to see that it trivially vanishes. Due to the antisymmetry of $L^{(1)}$, we have
\begin{eqnarray}
	\alpha_{\mu,ij}\big[L^{(1)}\big]^{ij}_{~k}&=&(\kappa_\mu P_{ij}+i\iota_\mu Q_{ij})\big[L^{(1)}\big]^{ij}_{~k}\\
	&=&i\iota_\mu Q_{ij}\big[L^{(1)}\big]^{ij}_{~k}.
\end{eqnarray}
Then \eqref{eq:linearterms} reads
\begin{eqnarray}
	 -\iota_\mu\iota_\nu~\left(Q_{ij}\big[L^{(1)}\big]^{ij}_{~k}\right)\left(Q_{i'j'}\big[L^{(1)}\big]^{i'j'}_{~k'}\right)\Omega^{kk'}=0
\end{eqnarray}
due to antisymmetry of $\Omega$. Line \eqref{eq:real} also vanishes identically due to the antisymmetry of $K$ and the symmetry of $(\kappa_\mu\kappa_\nu\PP_{ij}\PP_{i'j'}-\iota_\mu\iota_\nu \QQ_{ij}\QQ_{i'j'}){\mc N}^{ij}_{kl}{\mc N}^{i'j'}_{k'l'}$ under interchange of $(kl)\leftrightarrow(k'l')$. We are left with line \eqref{eq:imag}. Note that
\begin{eqnarray}\nonumber
&&\PP_{ij}\QQ_{i'j'}{\mc N}^{ij}_{kl}{\mc N}^{i'j'}_{k'l'}K^{klk'l'}\\
&&~~~~~~~~~~=\PP_{ij}\QQ_{i'j'}{\mc N_S}^{ij}_{kl}{\mc N_A}^{i'j'}_{k'l'}K^{klk'l'}
\label{eq:almostthere1}
\end{eqnarray}
where we have replaced $\mc N$ by their symmetric [$\mc N_S=\openone\otimes\openone$] and antisymmetric components [$\mc N_A=i(\tilde\Sigma\otimes\openone-\openone\otimes\tilde\Sigma)$] in accordance with the symmetry of $\PP$ and $\QQ$ respectively. Using Eq.~\eqref{eq:bicommutator} we can rewrite Eq.~\eqref{eq:almostthere1}
as
\begin{equation}
	\PP_{ij}\QQ_{i'j'}{\mc N}^{ij}_{kl}{\mc N}^{i'j'}_{k'l'}K^{klk'l'}=4\left(\tr[\QQ\tilde\Sigma
	\Sigma\PP\Omega]+\tr[\QQ\Sigma\PP\Sigma]\right)
\end{equation}
By virtue of $\tr[AB]=\tr[A^\top B^\top]$, $\QQ^\top=-\QQ$ and $(\Sigma\PP\Sigma)^\top=\Sigma\PP\Sigma$ the second term trivially vanishes. We are thus left with
\begin{equation}\label{eq:almostthere2}
	\tr[\rho[\Lambda_\mu,\Lambda_\nu]]=4i(\kappa_\mu\iota_\nu-\kappa_\nu\iota_\nu)\tr[\QQ \Sigma\Omega\Sigma\PP\Omega].
\end{equation}

In order to prove that this vanishes, let us analyze the structure of $\QQ$ and $\PP$. Let us express $\DD^{-1}[A]$ as
\begin{equation}\label{eq:summation}
	\DD^{-1}[A]_{ij}=-4\left[\sum_{k=0}^\infty4^k (\tilde \Sigma^\top)^k A \tilde \Sigma ^k\right]_{ij}.
\end{equation}
Given any matrix $A$ in block form
\begin{equation}
	A=\left(\begin{array}{cc}a_0 \openone & c_0 \ZZ \\c_0 \ZZ & b_0\openone\end{array}\right),
\end{equation}
we can define $A_k=(\tilde\Sigma^\top)^k A \tilde\Sigma^k$, and it is easy to show that
\begin{equation}
	\tilde\Sigma^\top A_k\tilde\Sigma=\left(\begin{array}{cc}a_{k+1}	\openone & c_{k+1} \ZZ \\c_{k+1} \ZZ &  b_{k+1}\openone\end{array}\right)
\end{equation}
with $a_{k+1}=a_k a^2+c (c b_k+2 a c_k)$, $b_{k+1}=a_k c^2+b (b b_k+2 c c_k)$ and $c_{k+1}=-c_k c^2-a a_k c-b b_k c-a b c_k$. From this it is straightforward to solve the recurrence equations and perform the resummation in Eq~\eqref{eq:summation} to obtain the explicit form of $\PP$. Similarly, one can obtain the explicit form for $\QQ$. For our purposes, however, it is only relevant to notice that
\begin{eqnarray}
	\PP=\left(\begin{array}{cc}a_p \openone & c_p \ZZ \\c_p \ZZ & b_p\openone\end{array}\right),\\
	\QQ=\left(\begin{array}{cc}a_q \omega & c_q \openone \\-c_q \openone & b_q\omega\end{array}\right).
\end{eqnarray}
Finally, we can obtain the structure of $\PP\Omega\QQ$, which reads
\begin{equation}
	\PP\Omega\QQ=\left(
	\begin{array}{cc}
		-(a_p a_q+c_pc_q)\openone&(a_pc_q-c_pb_q)\ZZ\\
		-c_p(a_q+c_q)\ZZ&c_p(c_q-b_q)\openone
	\end{array}
	\right).
\end{equation}
On the other hand we have
\begin{equation}
	\Sigma\Omega\Sigma=\left(
	\begin{array}{cc}
		(a^2-c^2)\omega&-c(a-b)\XX\\
		c(a-b)\XX&(b^2-c^2)\omega
\end{array}
\right),\quad \XX=\left(\begin{array}{cc}0&1\\1&0\end{array}\right).
\end{equation}
Finally, the trace in Eq.~\eqref{eq:almostthere2} reads
\begin{eqnarray}\nonumber
	\tr[\QQ \Sigma\Omega\Sigma\PP\Omega]&=&-(a_pa_q+c_pc_q)(a^2-c^2)\tr[\ZZ]\\
	\nonumber
	&&+(a_pc_q-c_pb_q)c(a-b)\tr[\ZZ \XX]\\
	\nonumber
	&&+c_p(a_q+c_q)c(a-b)\tr[\ZZ\XX]\\
	\nonumber
	&&+c_p(c_q-b_q)(b^2-c^2)\tr[\omega]\\
	&=&0.
\end{eqnarray}
This shows that
\begin{equation}
	\tr[\rho[\Lambda_\mu,\Lambda_\nu]]=0.
\end{equation}
To recapitulate, the expectation value of the commutator of the SLD operators vanishes for probe states given by the covariance matrix of Eq.~\eqref{eq:SVS}. In order to see that this result applies to all possible two-mode squeezed vacuum states, one has to observe that if $\Sigma'_0=S\Sigma_0 S^\top$, where $S$ are local phase shifts. Hence, the phase invariance of the channel guarantees that the output covariance matrix transforms in the same way, $\Sigma'=S\Sigma S^\top$, thus one can write
\begin{equation}
		 \tr[\rho[\Lambda_\mu,\Lambda_\nu]]=\alpha_{\mu,\,ij}\alpha_{\nu,\,kl}S^i_{~i'}S^j_{~j'}S^k_{~k'}S^l_{~l'}T^{i'j'k'l'},
\end{equation}
where the left hand side refers to the state corresponding to $\Sigma'$ and $T$ refers to the state $\Sigma$. Finally, since $S$ is a symplectic orthogonal transformation, and $\alpha_\mu$ is a linear combination of the symplectic and the Euclidean metrics, these are invariant under the congruence,
\begin{equation}
	\alpha_{\mu,\,ij}S^i_{~i'}S^j_{~j'}=\alpha_{\mu,\,i'j'}.
\end{equation}

This completes the proof of Eq.~\eqref{EQ:COMMUTE}, \emph{i.e.}, that the Heisenberg limit for $\gamma$ and $N$ corresponding to two-mode SVS can be, \emph{asymptotically}, simultaneously achieved.

\section{Exact expressions}\label{app:exact}
We report, for completeness, the exact expressions for the Fisher information for both parameters $\gamma$ and $N$.
When convenient, we will use the variables
\begin{eqnarray}
	x&=&n(n+1),\\
	y&=&N(N+1),\\
	z&=&e^\gamma-1.
\end{eqnarray}
Moreover, we find it useful to define $t=n+N+2nN=x+y-\Delta^2$ where $\Delta^2=(n-N)^2$. In terms of $x$ and $y$ we have
\begin{eqnarray}
	n&=&\frac{\sqrt{1+4x}-1}{2},\\
	N&=&\frac{\sqrt{1+4y}-1}{2},\\
	t&=&\frac{\sqrt{(1+4x)(1+4y)}-1}{2}.
\end{eqnarray}
The following inequalities are trivial consequences of the previous definitions,
\begin{subequations}
\begin{eqnarray}
	&x\geq&0,\\
	&y\geq&0,\\
	&z\geq&0,\\
	x+y\geq&t\geq&0.
\end{eqnarray}
\end{subequations}

Working out the results of~\cite{monras_information_2009} we obtain, for the different classes of states,	
\begin{subequations}
\begin{eqnarray}
\label{eq:exact.gamma.coherent}
	J_\gamma^\text{coh.}&=&\frac{N/z}{1+z(N+1)}+\frac{n}{1+z(2N+1)},
\\
	J_\gamma^\text{sq.}&=&\frac{t}{z}-\frac{y t}{t+y z}-\frac{2 (t (t+1)-y)}{(z+1)^2+2 z (t+y z)},~~~\\
	J_\gamma^\text{th.}&=&\frac{x+y-t}{y z^2+t z+x},\\
   J_\gamma^{\textrm{2-m}}&=&\frac{t+x z}{z ((t+1) z+1)}.
\end{eqnarray}
\end{subequations}
For $N$ we get
\begin{subequations}
\begin{eqnarray}
	J_N^\text{coh.}&=&\frac{z (z+1)^2}{(1+z(N+1))N},\\
	\nonumber
	J_N^\text{sq.}&=&\left[z (z+1)^2 (1+4 x+z (2 t+2 y z+z+2))\right]\\
	\nonumber
	&&\times\left[t+z(x (8 y+2)+y (z (2 y z+z+2)+3))\right.\\
	&&~~~~~\left.+t z^2 (1+4 y)\right],\\
	J_N^\text{th.}&=&\frac{z^2 (z+1)^2}{x+z (t+y z)},\\
   J_N^{\textrm{2-m}}&=&\frac{(t+1) z (z+1)^2}{y (t z+z+1)}.
\end{eqnarray}
\end{subequations}

It can be checked that
\begin{subequations}
\begin{eqnarray}\label{eq:thermineq}
	J_\gamma^{\textrm{2-m}}&\geq&J_\gamma^\text{th.},\\
	J_\gamma^{\textrm{2-m}}&\geq&J_\gamma^\text{coh.},\\
	J_\gamma^{\textrm{2-m}}&\geq&J_\gamma^\text{sq.}.
\end{eqnarray}
\end{subequations}
Although apparently nontrivial, a systematic method can be used to check all the above inequalities. We illustrate the procedure by considering Eq.~\eqref{eq:thermineq}. We first write it as
\begin{equation}
	\frac{t+x z}{z ((t+1) z+1)}\geq\frac{x+y-t}{y z^2+t z+x}
\end{equation}
to obtain
\begin{equation}
	(t+x z) \left(y z^2+t z+x\right)-z(x+y-t)((t+1) z+1)\geq 0.
\end{equation}
Now arranging in powers of $z$ we obtain
\begin{eqnarray}\nonumber
	x y z^3+\left(t(t+1)-x-y\right) z^2~~~~~~~~~~~~~~~~\\
	+\left(x^2+t(t+1)-x-y\right) z+t
   x\geq0.
\end{eqnarray}
Now it is trivial to show that all coefficients are positive by observing that $t(t+1)=x+y+4xy$. The same method applies to all inequalities. Notice that since $\gamma$ always appears as $\exp\gamma$, then only integer powers of $z$ will appear in the expressions. No series expansions will be necessary and therefore one only needs to check positivity for a finite number of coefficients.

\end{document}